\documentstyle[preprint,eqsecnum,aps]{revtex}

\begin{document}
\draft
\title{Neutron Scattering Studies of the Magnetic Fluctuations in 
YBa$_2$Cu$_3$O$_{7-\delta}$}
\author{H. A. Mook,$^1$ P. Dai,$^1$
R. D. Hunt,$^2$ and F. Do$\rm\breve{g}$an,$^3$}
\address{
$^1$Solid State Division,
 Oak Ridge National Laboratory\\
 Oak Ridge, 
Tennessee 37831-6393\\
$^2$Chemical Technology Division,
Oak Ridge National 
Laboratory\\
 Oak Ridge, Tennessee 37831-6221\\
$^3$Department of Materials Science and Engineering\\ 
University of Washington, Seattle, Washington 98195\\}
\maketitle
\begin{abstract}
Neutron scattering measurements have been made on the spin fluctuations 
in YBa$_2$Cu$_3$O$_{7-\delta}$ for different oxygen doping levels. 
Incommensurability is clearly observed for oxygen concentrations of 6.6 and 6.7 
and is suggested for the 6.93. Measurements of the resonance for the O$_{6.6}$ 
concentration show that it exists in a broadened and less intense form 
at temperatures much higher than $T_c$.   
\end{abstract}


\narrowtext
It now seems clear that the magnetic fluctuations  
play a central role in the behavior of high-$T_c$  
superconducting materials. Neutron scattering can determine both the 
wavevector and energy dependence of the magnetic fluctuations in a 
straightforward manner. However, the technique is strongly limited 
by the intensity of available neutron beams, thus requiring the use 
of large single crystal samples.  YBa$_2$Cu$_3$O$_{7-\delta}$  (YBCO) 
is one of the best characterized of the high-$T_c$ superconductors, 
but only now are crystals becoming available with sufficient quality 
and oxygen uniformity that reliable results can be obtained. This is 
particularly true of the underdoped materials that display 
interesting pseudogap behavior. In these materials very high oxygen 
uniformity is necessary, yet this is difficult with the large samples 
needed. We have now been successful in developing techniques that insure 
uniform oxygen concentrations through out the whole volume of large high quality crystals.

A second problem with the YBCO materials is that the 
energy scale of interest for the magnetic fluctuations is  
the same  as that for many of the intense phonon excitations.  
It is often very difficult to separate the magnetic excitations 
in neutron scattering experiments from the far more intense phonon excitations. 
We have approached the problem by using polarization analysis when 
possible to isolate the magnetic signal. In some cases, however, 
the magnetic scattering is too weak to permit the use of polarization 
analysis which is far less efficient that conventional techniques. 
Nevertheless, we have succeeded in better characterizing the phonon 
excitations through measurements and models so that a more reliable 
extraction of the magnetic signal is possible.

Another factor that has greatly aided our measurements is 
the development of the filter integration technique \cite{mook1}.  
In this technique use is made of the fact that the magnetic fluctuations are, 
except for the bilayer coupling, two-dimensional in nature and the scattering 
is summed over the $c^\ast$ direction in the reciprocal lattice 
greatly increasing the magnetic signal relative to other types of scattering. 
Graphite filters are used to essentially eliminate the elastic scattering 
so that only the fluctuations are measured. We do not obtain detailed 
energy selection with this technique, rather sample a band of energies. 
It is thus often desirable to perform standard measurements with triple-axis 
spectrometry to confirm some of the results if possible. However, the 
technique is excellent for survey studies and to measure very small signals. 

In this paper we will consider results on two of the most interesting aspects 
of the spin fluctuations in the YBCO materials, the resonance and the incommensurability. 
Let us first consider measurements for YBCO$_{6.6}$. 
The magnetic fluctuations are found around the reciprocal lattice position $(\pi, \pi)$ 
of the unit cell which is the position where antiferromagnetism is 
observed in the highly underdoped ordered compounds. This is the 
position $(0.5, 0.5, L)$ in reciprocal lattice units. We find the 
scattering in question stems from coupled bilayers so that $L$ 
is set for a value to sample acoustic scattering, in our case 1.7. 
To search for incommensurate fluctuations we scan along the line from 
$(0,0)$ to $(2\pi, 2\pi)$ or along $(H,H)$ in the reciprocal lattice.  
In the scan we use rather good resolution along the scan direction 
but must employ poor resolution perpendicular to this direction to obtain 
sufficient intensity. The scan direction is shown on the top of Fig 1. 
along with the position of possible incommensurate magnetic satellites. 
If the satellites  are as shown, a peak in the scattering will 
be found when the perpendicular to the scan direction intersects the 
left and lower satellites at the same time. A similar peak 
( smaller because of the magnetic form factor) will be found 
when the perpendicular intersects the upper and right hand 
satellites simultaneously. Another set of possible positions of 
the satellites would be along the scan direction in which case 
three peaks would be observed. Two from individual satellites on 
the scan direction and the third at the $(0.5,0.5)$ position as the 
perpendicular intersects peaks above and below the scan direction. 
The peak at $(0.5, 0.5)$ could be smaller depending on the 
extent of the resolution perpendicular to the scan. There could of 
course be other possible arrangements of the magnetic scattering. 

The graph in the lower part of Fig 1. shows a scan made for YBCO$_{6.6}$ 
using the filter integration technique. Peaks are observed on either side of the $(0.5,0.5)$ 
position for $(H,H)$ showing that the magnetic fluctuations are  
incommensurate. Further measurements with  triple axis  spectrometry 
show the fluctuations to be incommensurate \cite{dai1} at energies 
below the resonance energy of about 34 meV. The incommensurate position 
of the peaks from a series of scans was determined to be $(0.057, 0.057)$ with an error of 0.006. 
Note the actual satellite wavevector would be $2 \times 0.057$  
if the satellites are actually as pictured in Fig. 1. We are in the 
process of making sets of scans along different paths to determine the 
actual satellite wavevector.  This is much more difficult than for the 
214 superconductors because of the energy scale and the bilayer nature 
of the scattering.

We move now to measurements of the resonance scattering in YBCO$_{6.6}$. 
The resonance scattering in the fully doped YBCO materials is a very 
narrow peak found at 41 meV. The first observation of increased scattering at 
low temperatures at 41 meV in YBCO was made by Rossat-Mignod {\it et al.} \cite{mignod} 
although the low temperature response was considered a rather broad band of scattering 
at that time. Mook and coworkers \cite{mook2} showed using polarized beam 
techniques that for YBCO$_{6.93}$ the  41 meV scattering was a 
sharp peak that dominated the low temperature scattering. 
The measurements suggested that a small peak remained at 100 K. 
Fong {\it et al.} \cite{fong1} determined that for highly doped 
YBCO the resonance is not found above the superconducting transition temperature. 
The present results on YBCO$_{6.6}$ show that the resonance 
remains as a broadened and considerably less intense peak in energy 
at temperatures as high as a temperature $T^\ast$ which could be interpreted 
as the pseudogap temperature.

The resonance energy for YBCO$_{6.6}$ has been found to be about 34 meV \cite{dai2,fong2}. 
We show in Fig. 2 scans in energy through the resonance for six temperatures. 
These scans were obtained by making a series of scans in momentum q to determine 
the phonon background. The phonons tend to be reasonably flat in q near $(0.5, 1.5, 1.7)$ 
where the measurements were taken so they could be subtracted out with a 
considerable degree of confidence. Our scan in energy was then derived from 
the scans in momentum. Our success at removing the phonon background is 
verified by the 293K data which shows little scattering as expected for 
the magnetic excitations.

The resonance at 10 K is resolution limited for our experiment. 
As the temperature is raised through $T_c=63$ K the resonance broadens and 
diminishes in intensity, but is still observed as a definite peak in energy.  
A signature of the resonance remains at 150 K , but appears to vanish by about 200 K 
with little scattering observed by 293K. A scan of intensity vs. temperature 
made at the resonance energy of 34 meV shown in Ref. \cite{dai2} indicates that 
the character of the scattering changes at $T_c$. 
The scattering intensity has a noticeable upturn upon cooling through $T_c$ and 
increases more rapidly as the temperature is lowered further. A possible way 
to consider the observed behavior is to characterize the resonance as a property 
of the superconducting state with a precursor that exists in the normal state 
at temperatures as high as a pseudogap temperature $T^\ast$. 

We now turn to measurements on a sample with an oxygen concentration of 6.7, 
YBCO$_{6.7}$. We find that $T_c$ is 74 K for this sample while the resonance energy is 38.5 meV. 
Fig. 3 shows our measurements of the magnetic scattering  made for this sample 
using the filter integration technique. The scan direction is along $(\pi, \pi)$ 
and was done in the same way as the scan shown in Fig. 1.  Again we observe the 
scattering is incommensurate at low temperatures. The incommensurability is still 
observed as the temperature is raised to 100 K, but is clearly gone by 200 K. 
Little if any scattering is observed at room temperature showing that the 
scans are free of phonon contributions. Least squares analysis of low temperature 
scans show an incommensurability wavevector of $(0.061, 0.061)$ with an error of 0.005. 
The measured incommensurate wavevector is thus larger than for the YBCO6.6 sample, 
but the error in position is sufficiently large that a clear difference 
is not established. The precise determination of the incommensurability wavevector 
is hampered by the weakness of the scattering coupled with an energy scale that 
results in reduced momentum resolution.  

We have performed preliminary experiments to search for inconmensurate 
fluctuations in the optimally doped material YBCO$_{6.93}$. 
We use the integration technique once again and the results are shown in Fig. 4. 
The integration technique  is not as favorable in this case as the energy window 
passed by the filter now includes the resonance to some degree. The energy 
sensitivity of the experiment is given in Ref. \cite{dai1}. The scan at 10 K shown 
in the top graph is thus expected to show a peak at $(H,H) = 0.5$ 
that stems from the resonance. We see in addition unresolved structure 
surrounding the peak at 0.5 that can be interpreted as stemming from 
incommensurate fluctuations. This structure largely disappears at 100 K with the 
peak at  0.5 now likely stemming from normal state commensurate excitations. 
Since now $T^\ast$ is much closer to $T_c$ for the O$_{6.93}$ material 
the 100 K spectrum appears similar to the 200 K spectrum for the O$_{6.7}$ sample. 
The counting rate for the O$_{6.93}$ experiment is considerably higher than 
for the other concentrations as a much bigger sample was used and the collimation 
was relaxed to some degree. The bottom panel of Fig. 4 shows the result of taking 
the difference between the 10 K and 100 K measurements. This result gives 
an indication of the nature of the incommensurate part of the scattering pattern. 
The intensity of the magnetic scattering decreases rapidly as higher oxygen 
doping levels are used so the counting errors, particularly in the difference pattern, 
are larger than desirable. It can also be misleading to use the 
differences of patterns taken at different temperatures to determine lineshapes. 
Nevertheless, the difference pattern serves as an indication of possible 
incommensurability in optimally doped YBCO. A least squares fit of 
Gaussian lineshapes to the pattern shown in the figure results in 
an incommensurability wavevector of $(0.08, 0.08)$. 
If the satellites are indeed in the position shown in Fig. 1 the 
incommensurability wavevector would be $2\times0.08$ or a value near the wavevector 
for the optimally doped 214 material. However, additional measurements are needed 
to confirm the  incommensurability of YBCO$_{6.93}$ and 
to determine the momentum space positions of satellite peaks.

We will not discuss in this paper how these results affect various theories 
of high $T_c$ superconductivity. However, we will make some general observations. 
It is important that the resonance not only shifts it's position in the 
underdoped materials, but  that its basic character is modified. In the 
heavily doped material the resonance begins to appear only near Tc while 
precursor effects with a temperature scale that matches the pseudogap 
temperature are observed in the underdoped case.  We note the resonance 
is affected by the bulk superconducting transition as a change in slope 
in the temperature dependence occurs at $T_c$. 
Theories to explain the resonance must take these observations into consideration. 

The observation of incommensurability is important as it unifies 
the picture of magnetic fluctuations in the high-$T_c$ materials. The 
preliminary evidence we have on the doping dependence suggests a close connection 
between the 214 and 123 materials. Additional measurements are needed to make 
detailed connections. Obviously any theoretical models that can account for the 
incommensurability would now appear to be applicable to the high-$T_c$ 
materials in general.  This research was supported by the U.S. DOE under 
contract No. DE-AC05-96OR22464 with Lockheed Martin Energy Research, Inc.

\begin{figure}
\caption{The top panel shows a possible position of the 
incommensurate satellites in YBCO. A scan along the arrow would 
produce two peaks stemming from the very relaxed resolution perpendicular 
to the scan direction. The bottom graph shows such a scan for YBCO$_{6.6}$.}
\label{autonum}
\end{figure}

\begin{figure}
\caption{Triple-axis scans through the resonance energy 
for YBCO$_{6.6}$. We note the resonance peak survives 
in a broadened and less intense form at temperatures as high as 150 K.}
\end{figure}

\begin{figure}
\caption{Scans made along the direction shown in Fig. 1 for YBCO$_{6.7}$. 
Incommensurability is observed at low temperatures with little magnetic 
intensity remaining at 300 K.}
\end{figure}

\begin{figure}
\caption{Scans for YBCO$_{6.93}$ made in the same way as for the 
underdoped materials except that a low angle 
background determined by rotating the crystal has been substracted. 
Structure in the 10 K scan suggests incommensurability 
that is more clearly shown in the difference result shown in the bottom panel.             }
\end{figure}


\begin{references}
\bibitem{mook1} H. A. Mook {\it et al.}, Phys. Rev. Lett. {\bf 77}, 370 (1996).
\bibitem{dai1} P. Dai, H. A. Mook, and F. Do$\rm\breve{g}$an, cond-mat/9707112.
\bibitem{mignod} J. Rossat-Mignod {\it et al.}, Physica (Amsterdam) {\bf 185C}, 86 (1991).
\bibitem{mook2} H. A. Mook {\it et al.}, Phys. Rev. Lett. {\bf 70}, 3490 (1993).
\bibitem{fong1} H. F. Fong {\it et al.}, Phys. Rev. Lett. {\bf 75}, 316 (1995).
\bibitem{dai2} P. Dai {\it et al.}, Phys. Rev. Lett. {\bf 77}, 5425 (1996).
\bibitem{fong2} H. F. Fong {\it et al.}, Phys. Rev. Lett. {\bf 78}, 713 (1997).
\end{references}
\end{document}